\documentclass[times,twocolumn,final]{article}

\usepackage{caption}
\usepackage[utf8x]{inputenc}
\usepackage{authblk}

\usepackage{algorithm}
\usepackage{algorithmic}
\usepackage{supertabular} 

\usepackage{graphicx}
\usepackage[colorinlistoftodos]{todonotes}
\usepackage[colorlinks=true]{hyperref}
\usepackage{multirow}
\usepackage{blindtext}
\usepackage{graphicx}
\usepackage{csvsimple}
\usepackage{hyperref}
\usepackage{graphics}
\usepackage{todonotes}
\usepackage{tikz}
\usepackage{amssymb}
\usepackage{subfigure}
\usepackage{amsmath}

\usepackage{array}
\usepackage{enumitem}
\usepackage{amsthm}
\usepackage{natbib}
\usepackage{vmargin}
\usepackage{minitoc}
\usepackage{rotating}
\usepackage{graphicx}
\usepackage{amssymb}

\newcounter{para}

\newtheorem{proposition}{Proposition}
\newtheorem{property}{Property}

\newtheorem{definition}{Definition}
\newtheorem{problem}{Problem}
\newtheorem{model}{Model}

\providecommand{\keywords}[1]
{
  \small	
  \textbf{\textit{Keywords---}} #1
}

\begin{document}



\title{On the unification of the graph edit distance and graph matching problems}



\author[1]{Romain Raveaux}

\date{7th March 2020}
\affil[1]{Universit\'{e} de Tours, Laboratoire d'Informatique Fondamentale et Appliqu\'{e}e de Tours (LIFAT - EA 6300), 64 Avenue Jean Portalis, 37000 Tours, France}


\onecolumn
\maketitle

\begin{abstract}
Error-tolerant graph matching gathers an important family of problems. These problems aim at finding correspondences between two graphs while integrating an error model. In the Graph Edit Distance (GED) problem, the insertion/deletion of edges/nodes from one graph to another is explicitly expressed by the error model. At the opposite, the problem commonly referred to as ``graph matching" does not explicitly express such operations. For decades, these two problems have split the research community in two separated parts. It resulted in the design of different solvers for the two problems.
In this paper, we propose a unification of both problems thanks to a single model. We give the proof that the two problems are equivalent under a reformulation of the error models. This unification makes possible the use on both problems of existing solving methods from the two communities.

\end{abstract}


\begin{figure*}
\begin{center}
\includegraphics[scale=0.5]{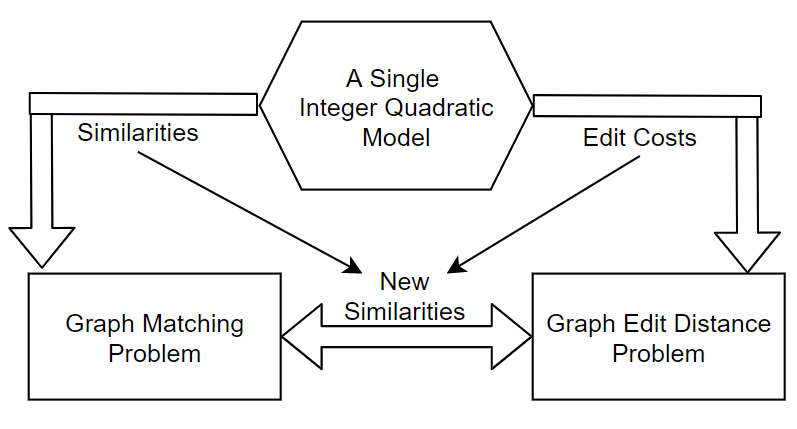}
\caption*{Graphical Abstract}
\end{center}
\end{figure*}


\twocolumn

\section{Introduction}
Graphs are frequently used in various fields of computer science, since they constitute a universal modeling tool which allows the description of structured data. The handled objects and their relations are described in a single and human-readable formalism. Hence, tools for graphs supervised classification and graph mining are required in many applications such as pattern recognition \citep{Riesen2015bouquin}, chemical components analysis, transfer learning \citep{DAS201880}. In such applications, comparing graphs is of first interest. The similarity or dissimilarity between two graphs requires the computation and the evaluation of the ``best" matching between them. Since exact isomorphism rarely occurs in pattern analysis applications, the matching process must be error-tolerant, i.e., it must tolerate differences on the topology and/or its labeling. The Graph Edit Distance (GED)\citep{Riesen2015bouquin} problem and the Graph Matching problem (GM) \citep{Swoboda_2017_CVPR} provide two different error models. These two problems have been deeply studied but they have split the research community into two groups of people developing separately quite different methods. 

In this paper, we propose to unify the GED problem and the GM problem in order to unify the work force in terms of methods and benchmarks. We show that the GED problem can be equivalent to the GM problem under certain (permissive) conditions. The paper is organized as follows: Section 2, we give the definitions of the problems. Section 3, the state of the art on GM and GED is presented along with the literature comparing GED and GM to other problems. Section 4, a specific related work is detailed since it is the basement of our reasoning. Section 5, our proposal is described and a proof is given. Section 6, experimental results are presented to validate empirically our proposal. Finally, conclusions are drawn.

\section{Definitions and problems}
In this section, we define the problems to be studied.
An attributed graph is considered as a set of 4 tuples ($V$,$E$,$\mu$,$\zeta$) such that: $G$ = ($V$,$E$,$\mu$,$\zeta$). $V$ is a set of vertices. $E$ is a set of edges such as $E \subseteq V \times V$. $\mu$ is a vertex labeling function which associates a label to a vertex. $\zeta$ is an edge labeling function which associates a label to an edge.


\subsection{Graph matching problem}
The objective of graph matching is to find correspondences between two attributed graphs $G_1$ and $G_2$. A solution of graph matching is defined as a subset of possible
correspondences $\mathcal{Y} \subseteq V_1 \times V_2$, represented by a binary assignment matrix $Y \in \{0,1 \}^{n1 \times n2}$, where $n1$ and $n2$ denote the number of nodes in $G_1$ and $G_2$, respectively. If $u_i \in V_1$ matches $v_k \in V_2$, then $Y_{i,k}=1$, and $Y_{i,k}=0$ otherwise. We
denote by $y\in \{0,1 \}^{n1 . n2}$, a column-wise vectorized replica of $Y$. With this notation, graph matching problems can be expressed as finding the assignment vector $y^*$ that maximizes a score function $S(G_1, G_2, y)$ as follows:
\begin{model}{Graph matching model (GMM)}
\label{model:iqpsubgm}
\begin{subequations}
  \begin{align}
    y^* =& \underset{y} {\mathrm{argmax}} \quad S(G_1, G_2, y)\\
    \text{subject to}\quad & y_{i,k} \in \{0, 1\} \quad \forall (u_i, v_k) \in V_1 \times V_2\\
    \label{eq:subgmgmc}
    &\sum_{u_i \in V_1} y_{i,k} \leq 1 \quad \forall v_k \in V_2\\
    \label{eq:subgmgmd}
     &\sum_{v_k \in V_2} y_{i,k} \leq 1 \quad \forall u_i \in V_1
  \end{align}
\end{subequations}
\end{model}
where equations \eqref{eq:subgmgmc},\eqref{eq:subgmgmd} induces the matching constraints, thus making $y$ an assignment vector.

The function $S(G_1, G_2, y)$ measures the similarity of graph attributes, and is typically decomposed into a first order similarity function $s(u_i \to v_k)$ for a node pair $u_{i} \in V_1$ and $v_k \in V_2$, and a second-order similarity function $s(e_{ij} \to e_{kl})$ for an edge pair $e_{ij} \in E_1$ and $e_{kl} \in E_2$. Thus, the objective function of graph matching is defined as: 
  \begin{equation}
  \label{eq:matchingfunction}
  \begin{aligned}
  S(G_1,G_2,y) =&	\sum_{u_i \in V_1}\sum_{v_k \in V_2} s(u_i \to v_k)  \cdot y_{i,k} \\ 
  &+ \sum_{e_{ij} \in E_1}\sum_{e_{kl} \in E_2} s(e_{ij} \to e_{kl}) \cdot  y_{ik} \cdot y_{jl}
    \end{aligned}
\end{equation}
 
In essence, the score accumulates all the similarity values that are relevant to the assignment. \color{black}The GM problem has been proven to be $\mathcal{NP}$-hard by \citep{GMcomplexity}.\color{black}

\subsection{Graph Edit Distance}

\label{subsec:gedproblem}
The graph edit distance (GED) was first reported in \citep{Tsai1979}. GED is a dissimilarity measure for graphs that represents the minimum-cost sequence of basic editing operations to transform a graph into another graph by means classically included operations: insertion, deletion and substitution of vertices and/or edges. Therefore, GED can be formally represented by the minimum cost edit path transforming one graph into another. Edge operations are taken into account in the matching process when substituting, deleting or inserting their adjacent vertices. From now on and for simplicity, we denote the substitution of two vertices $u_i$ and $v_k$ by ($u_i \to v_k$), the deletion of vertex $u_i$ by ($u_i \to \epsilon $) and the insertion of vertex $v_k$ by ($ \epsilon \to v_k$). Likewise for edges $e_{ij}$ and $e_{kl}$, ($e_{ij} \to e_{kl}$) denotes edges substitution, ($e_{ij} \to \epsilon$) and ($\epsilon \to e_{kl}$) denote edges deletion and insertion, respectively. 


An edit path ($\lambda$) is a set of edit operations $o$. This set is referred to as \emph{Edit Path} and it is defined in Definition \ref{chap2:editpath}.
\begin{definition}{Edit Path}\\
A set $\lambda=\{o_1, \cdots , o_k\}$ of $k$ edit operations $o$ that transform $G_1$ completely into $G_2$ is called an (complete) edit path.
\label{chap2:editpath}
\end{definition}
Let $c(o)$ be the cost function measuring the strength of an edit operation $o$. Let $\Gamma(G_1,G_2)$ be the set of all possible edit paths ($\lambda$).
The graph edit distance problem is defined by Problem \ref{prob:ged}.

\begin{problem}{Graph Edit Distance}\\
Let  $G_1$ = ($V_1$,$E_1$,$\mu_{1}$,$\zeta_{1}$) and $G_2$ = ($V_2$,$E_2$,$\mu_{2}$,$\zeta_{2}$) be two graphs, the graph edit distance between $G_1$ and $G_2$ is defined as:

\begin{equation}
d_{min}(G_1,G_2) = \min_{\lambda \in \Gamma(G_1,G_2)} \sum_{o \in \lambda }{c(o)}
\end{equation}
\label{prob:ged}
\end{problem}
The GED problem is a minimization problem and $d_{min}$ is the best distance.
In its general form, the GED problem (Problem \ref{prob:ged}) is very versatile. The problem has to be refined to cope with the constraints of an assignment problem.
First, let us define constraints on edit operations $(o_i)$ in Definition \ref{def:operationconstraint}.
\begin{definition}{Edit operations constraints }
\begin{enumerate}
\item Deleting a vertex implies deleting all its incident edges.
\item Inserting an edge is possible only if the two vertices already exist or have been inserted.
\item Inserting an edge must not create more than one edge between two vertices. 
\end{enumerate}
\label{def:operationconstraint}
\end{definition}
Second, let us define constraints on edit paths $(\lambda$) in Definition \ref{def:editpathconstraint}. This type of constraint prevents the edit path to be composed of an infinite number of edit operations.
\begin{definition}{Edit path constraints }
\begin{enumerate}
\item $k$ is a finite positive integer.
\item A vertex/edge can have at most one edit operation applied on it.
\end{enumerate}
\label{def:editpathconstraint}
\end{definition}
Finally, let us define the topological constraint in Definition \ref{def:topologyconstraint}. This type of constraints avoids edges to be matched without respect to their adjacent vertices.
\begin{definition}{Topological constraint }\\
The topological constraint implies that matching (substituting) two edges $(u_i , u_j) \in E_1$ and $(v_k , v_l) \in E_2$ is valid if and only if their incident vertices are matched $(u_i \to v_k)$ and $(u_j \to v_l)$. 
\label{def:topologyconstraint}
\end{definition}

An important property of the GED can be inferred from the topological constraint defined in Definition \ref{def:topologyconstraint}.
\begin{property}{The edges matching are driven by the vertices matching}

Assuming that constraint defined in Definition \ref{def:topologyconstraint} is satisfied then three cases can appear : \\
Case 1: If there is an edge $e_{ij}$ = $(u_i,u_j) \in E_1$ and an edge $e_{kl}$ = $(v_k,v_l) \in E_2$, edges substitution  between  $(u_i,u_j)$ and $(v_k,v_l)$ is performed (i.e., ($e_{ij} \to e_{kl}$)). 

\noindent Case 2: If there is an edge $e_{ij}$ = $(u_i,u_j) \in E_1$ and there is no edge between $v_k$ and $v_l$ then an edge deletion of $(u_i,u_j)$ is performed (i.e., ($e_{ij} \to \epsilon $)). \\
\noindent Case 3: If there is no edge  between $u_i$ and $u_j$ and there is an edge between and an edge $e_{kl}$ = $(v_k,v_l) \in E_2$ then an edge insertion of $(v_k,v_l)$ is performed (i.e., ($ \epsilon  \to e_{kl} $)).
\label{def:gedproperty}
\end{property}

The GED problem defined in Problem \ref{prob:ged} and refined with constraints defined in Definitions \ref{def:operationconstraint}, \ref{def:editpathconstraint} and \ref{def:topologyconstraint} is referred in the literature and in this paper as the GED problem. \color{black}The GED problem has been proven to be $\mathcal{NP}$-hard by \citep{DBLP:journals/pvldb/ZengTWFZ09}.\color{black} 

\subsection{Related problems and models}
GED and GM problems fall into the family of error-tolerant graph matching problems. GED and GM problems can be equivalent to another problem called Quadratic Assignment Problem (QAP) \citep{bougleuxQAP,DBLP:conf/iccv/ChoAP13}. In addition, GED and GM problems can be equivalent to a constrained version of the Maximum a posteriori (MAP)-inference problem of a Conditional Random Field (CRF) \citep{Swoboda_2017_CVPR}. All these problems can be expressed by mathematical models. A mathematical model is composed of variables, constraints and an objective functions. A single problem can be expressed by many different models. An Integer Quadratic Program (IQP) is a model with a quadratic objective function of the variables and linear constraints of the variables. We chose to present the GM problem as an IQP (Model \ref{model:iqpsubgm}). At the opposite, an Integer Linear Program (ILP) is a mathematical model where the objective function is a linear combination of the variables. The objective function is constrained by linear combinations of the variables.

\section{State of the art}
In this section, the state of the art is presented. First, the solution methods for GED and GM are described. Finally, papers comparing GED to other matching problems are mentioned.
\subsection{State of the art on GM and GED}
\color{black}The GED and GM problems have been proven to be $\mathcal{NP}$-hard. So, unless  $\mathcal{P} = \mathcal{NP}$, solving the problem to optimality cannot be done in polynomial time of the size of the input graphs. Consequently, the runtime complexity of exact methods is not polynomial but exponential with respect to the number of vertices of the graphs. On the other hand, heuristics are used when the demand for low computational time dominates the need to obtain optimality guarantees.\color{black}
\paragraph{GM methods} 
 Many solver paradigms were put to the test for GM. These include relaxations based on Lagrangean decompositions \citep{Swoboda_2017_CVPR,messagepassingdualdecomposition}, convex/concave quadratic \citep{gnccp} (GNCCP) and semi-definite programming \citep{pbmatching} , which can be used either directly to obtain approximate solutions or just to provide lower bounds. To tighten these bounds several cutting plane methods were proposed \citep{Bazaraa1982}. On the other side, various primal heuristics, both (i) deterministic, such as graduated assignment methods \citep{graduatedassignment}, fixed point iterations \citep{IPFP} (IPFP), spectral technique and its derivatives \citep{SMACGM,spectralmatching} and (ii) non-deterministic (stochastic), like random walk \citep{reweightedgm} were proposed to provide approximate solutions to the problem. Many of these methods were published in TPAMI, NIPS, CVPR, ICCV.
\paragraph{GED methods} Exact GED algorithms were proposed based on tree search \citep{Tsai1979,DBLP:conf/mlg/RiesenFB07,DBLP:conf/icpram/Abu-AishehRRM15}. Another way to build exact methods is to model the problem by Integer Linear Programs. Then, a black box solver is used to obtain solutions \citep{justiceheroged2006,DBLP:journals/pr/LerougeARHA17}. In addition, the GED community worked on simplifications of the GED problem to the Linear Sum Assignment Problem (LSAP) \citep{bougleuxlsap,DBLP:journals/ivc/Serratosa15,DBLP:journals/ivc/RiesenB09}. The GED problem was modeled as a QAP \citep{bougleuxQAP}.  \color{black} Let us named this model \textbf{GEDQAP}. The GEDQAP model has extra variables to cope with the insertion and deletions cases and all costs are represented by a $(|V_1| + |V_2|)^2 \times (|V_1| + |V_2|)^2$ matrix $D$.
The cost matrix $D$ can be decomposed as follows into four blocks of size $(|V_1| + |V_2|) \times (|V_1| + |V_2|)$.
The left upper block of the matrix $D$ contains all possible edge substitutions, the diagonal of the right upper matrix represents the cost of all possible edge deletions and the diagonal of the bottom left corner contains all possible edge insertions. Finally, the bottom right block elements cost is set to a large constant $w$ which concerns the matching of $\epsilon - \epsilon$ edges.
The GEDQAP model has $ (|V_1| + |V_2|)^2 $ variables and $ (|V_1| + |V_2|) + (|V_1| + |V_2|) $ constraints. The cost matrix size is $ (|V_1| + |V_2|)^2 \times (|V_1| + |V_2|)^2$.  \color{black} Based on this GEDQAP model, modified versions of IPFP \citep{bougleuxQAP} and GNCCP \citep{bougleuxQAP} were proposed. Finally, many GED methods were published in PRL, PR, Image and Vision Computing, GbR, SSPR.

\subsection{State of the art on comparing GED problems to others}
Neuhaus and Bunke \citep{Neuhaus2007} have shown that if each operation cost satisfies the criteria of a distance (positivity, uniqueness, symmetry, triangular inequality) then the edit distance defines a metric between graphs and it can be inferred that if $GED(G_1,G_2)=0 \Leftrightarrow G_1 = G_2$. Furthermore, it has been shown that standard concepts from graph theory, such as graph isomorphism, subgraph isomorphism, and maximum common subgraph, are special cases of the GED problem under particular cost functions \citep{DBLP:journals/prl/Bunke97,DBLP:journals/pami/Bunke99,brun:hal-00714879}. 

\paragraph*{Deadlocks, contributions and motivations}
From the literature, two main deadlocks can be drawn. First, GED and GM problems split the research community in two parts. People working on GED do not work on GM and vice and versa. They do not contribute to the same journals and conferences. Second, these two communities do not use the same methods to solve their problem while they have mainly the same applications fields (computer vision, chemoinformatics, ...). Researchers working on GM problems have concentrated their efforts on the QAP and MAP-inference solvers (Frank-Wolfe like methodology \citep{IPFP,gnccp}, Lagrangian decomposition methods \citep{Swoboda_2017_CVPR,messagepassingdualdecomposition}, ...). On the other hand, the community working on the GED problem has favored LSAP-based and tree-based methods.

Our motivation is to gather people working on GED and GM problems because methods and benchmarks built from one community could help the other. A first step forward has been done by \citep{bougleuxQAP} by modelling the GED problem as a specific QAP and using modified solvers from the graph matching community. \color{black}However, our proposal stands apart from their work because we propose a \textbf{single model} to express the  \textbf{GM} and the \textbf{GED} problems. \color{black}
In this direction, we propose more investigations to compare GED and GM problems. We propose a theoretical study to relate GM and GED problems. Our contribution is to prove that GED and GM problems are equivalent in terms of solutions under a
reformulation of the similarity function. Consequently, all the methods solving the GM problem can be used to solve the GED problems.


\section{Related works: Integer Linear Program for GED}
In \citep{DBLP:journals/pr/LerougeARHA17}, an ILP was proposed to model the GED problem. This model will play an important role in our proposal so we propose to give a brief definition of this model. For each type of edit operation, a set of corresponding binary variables is defined in Table \ref{tab:ILPvar}.

\begin{table}[]
\begin{scriptsize}
    \centering
    \caption{Definition of binary variables of the ILP.}
    \begin{tabular}{|c||c|p{2.5cm}|}
    \hline
         Name&Card&Role  \\\hline\hline
         $y_{i,k}$& $\forall (u_i,v_k) \in V_1 \times V_2$&=1 if $u_i$ is substituted with $v_k$\\\hline
         $ z_{ij,kl}$& $\forall (e_{ij},e_{kl}) \in E_1 \times E_2$&=1 if $e_{ij}$ is substituted with $e_{kl}$\\\hline
         $ a_i$& $\forall u_i \in V_1$&=1 if $u_i$ is deleted from $G_1$ \\\hline
         $ b_{ij}$& $\forall e_{ij} \in E_1$&=1 if $e_{ij}$ is deleted from $G_1$ \\\hline
          $g_k$& $\forall v_k \in V_2$&=1 if $v_k$ is inserted in $G_1$ \\\hline
         $ h_{kl}$& $\forall e_{kl} \in E_2$&=1 if $e_{kl}$ is inserted in $G_1$ \\\hline
    \end{tabular}
    \label{tab:ILPvar}
\end{scriptsize}
\end{table}

The objective function \eqref{eq:objective} is the overall cost induced by an edit path $({y}, {z}, {a}, {b}, {g}, {h})$ that transforms a graph $G_{1}$ into a graph $G_{2}$. In order to get the graph edit distance between $G_{1}$ and $G_{2}$, this objective function must be minimized.
\begin{equation}
\begin{aligned}
  &C({y,z,a,b,g,h}) = \Biggl( \sum_{u_i \in V_1}\sum_{v_k \in V_2} c(u_i \rightarrow v_k) \cdot y_{i,k}\\
  & + \sum_{e_{ij} \in E_1}\sum_{e_{kl} \in E_2} c(e_{ij} \rightarrow e_{kl}) \cdot z_{ij,kl} 
  + \sum_{u_i \in V_1} c(u_i \rightarrow \epsilon) \cdot a_i \\
  &+ \sum_{v_k \in V_2} c(\epsilon \rightarrow v_k) \cdot g_k 
  + \sum_{e_{ij} \in E_1} c(e_{ij} \rightarrow \epsilon) \cdot b_{ij} \\
  &+ \sum_{e_{kl} \in E_2} c(\epsilon \rightarrow e_{kl}) \cdot h_{kl} \Biggr)
\end{aligned}
\label{eq:objective}
\end{equation}

Now, the constraints are presented. They are mandatory to guarantee that the admissible solutions of the ILP are edit paths that transform $G_{1}$ in $G_{2}$. 
The constraint \eqref{eq:sumx1} ensures that each vertex of $G_{1}$ is either mapped to exactly one vertex of $G_{2}$ or deleted from $G_{1}$, while the constraint \eqref{eq:sumx2} ensures that each vertex of $G_{2}$ is either mapped to exactly one vertex of $G_{1}$ or inserted in $G_{1}$:
\begin{subequations}
  \begin{align}
  a_i + \sum_{v_k \in V_2} y_{i,k}  = 1 \quad \forall u_i \in V_1 \label{eq:sumx1}\\
  g_k + \sum_{u_i \in V_1} y_{i,k}  = 1 \quad \forall v_k \in V_2 \label{eq:sumx2}
  \end{align}
\end{subequations}
The same applies for edges:
\begin{subequations}
  \begin{align}
    b_{ij} + \sum_{e_{kl} \in E_2} z_{ij,kl}  = 1 \quad \forall e_{ij} \in E_1 \label{eq:sumy1}\\
  h_{kl} + \sum_{e_{ij} \in E_1} z_{ij,kl}  = 1 \quad \forall e_{kl} \in E_2 \label{eq:sumy2}
  \end{align}
\end{subequations}


The topological constraints defined in Definition \ref{def:topologyconstraint} can be expressed with the following constraints \eqref{eq:topology_1} and \eqref{eq:topology_2}:

$e_{ij}$ and $e_{kl}$ can be mapped only if their head vertices are mapped:
	\begin{equation}
	  z_{ij,kl} \leq y_{i,k} \quad \forall (e_{ij}, e_{kl}) \in E_1 \times E_2
	  \label{eq:topology_1}
	\end{equation}
 $e_{ij}$ and $e_{kl}$ can be mapped only if their tail vertices are mapped:
	\begin{equation}
	  z_{ij,kl} \leq y_{j,l} \quad \forall  (e_{ij}, e_{kl}) \in E_1 \times E_2
	  \label{eq:topology_2}
	\end{equation}

 The insertions and deletions variables $a$, $b$, $g$ and $h$ help the reader to understand how the objective function and the constraints were obtained, but they are unnecessary to solve the GED problem. In the equation \eqref{eq:objective}, the variables ${a}, {b}, {g} \text{ and } {h}$ are replaced by their expressions deduced from the equations \eqref{eq:sumx1}, \eqref{eq:sumx2}, \eqref{eq:sumy1} and \eqref{eq:sumy2}. For instance, from the equation \eqref{eq:sumx1}, the variable $a$ is deduced: $a_i=1-\sum_{v_k \in V_2} y_{i,k} $ and replaced in the equation \eqref{eq:objective}, the part of the objective function concerned by variable $a$ becomes: 
 \begin{equation}
 \begin{aligned}
  \sum_{u_i \in V_1} c(u_i \rightarrow \epsilon) \cdot a_i=&
  \sum_{u_i \in V_1} c(u_i \rightarrow \epsilon) \\
  &+\sum_{u_i \in V_1}\sum_{v_k \in V_2} -c(u_i \rightarrow \epsilon).y_{i,k} \\
  \end{aligned}
 \end{equation}
 Consequently, a new objective function is expressed as follows:
\begin{equation}
 \begin{aligned}
  C'({y,z})=&  \gamma +
        \sum_{u_i \in V_1}\sum_{v_k \in V_2}
        \Bigl(c(u_i \rightarrow v_k) \\
        &- c(u_i \rightarrow \epsilon)
        - c(\epsilon \rightarrow v_k)\Bigr) \cdot y_{i,k} \\
        &+ \sum_{e_{ij} \in E_1}\sum_{e_{kl} \in E_2} \Bigl(c(e_{ij} \rightarrow
        e_{kl})\\
        &- c(e_{ij} \rightarrow \epsilon)
        - c(\epsilon \rightarrow e_{kl})\Bigr) \cdot z_{ij,kl} \\
\text{with } \gamma =  & \sum_{u_i \in V_1} c(u_i \to \epsilon)+ \sum_{v_k \in V_2} c(\epsilon \to v_k) \\
  & + \sum_{e_{ij} \in E_1} c(e_{ij} \to \epsilon) + \sum_{e_{kl} \in E_2} c(\epsilon \to e_{kl}) 
\end{aligned}
\label{eq:objective2}
\end{equation}
Equation \eqref{eq:objective2} shows that the GED can be obtained without explicitly computing the variables ${a}, {b}, {g} \text{ and } {h}$. Once the formulation solved, all insertion and deletion variables can be \emph{a posteriori} deduced from the substitution variables. 

The vertex mapping constraints \eqref{eq:sumx1} and \eqref{eq:sumx2} are transformed into inequality constraints, without changing their role in the program. As a side effect, it removes the $a$ and $g$ variables
from the constraints:
\begin{equation} \label{eq:sumx1b}
  \sum_{v_k \in V_2} y_{i,k} \leq 1 \quad \forall u_i \in V_1
\end{equation}
\begin{equation} \label{eq:sumx2b}
  \sum_{u_i \in V_1} y_{i,k} \leq 1 \quad \forall v_k \in V_2
\end{equation}




\color{black}In fact, the insertions and deletions variables $a$ and $g$ of the equations \eqref{eq:sumx1} and \eqref{eq:sumx2} can be seen as slack variables to transform inequality constraints to equalities and consequently providing a \textit{canonical form}. \color{black}The entire formulation is called F2 and described as follows : 
\begin{model}{F2}
\begin{subequations}
    \begin{align}
        &\min_{{y,z}} C'(y,z) \label{f2:o}\\
    \text{subject to}\quad
    &\sum_{v_k \in V_2} y_{i,k} \leq 1 \quad \forall u_i \in V_1\label{f2:c1}\\
    &\sum_{u_i \in V_1} y_{i,k} \leq 1 \quad \forall v_k \in V_2\label{f2:c2}\\
    &\sum_{e_{kl} \in E_2} z_{ij,kl} \leq y_{i,k} \quad \forall v_k \in V_2, \forall e_{ij} \in E_1\label{f2:c3}\\
    &\sum_{e_{kl} \in E_2} z_{ij,kl} \leq y_{j,l} \quad \forall v_l \in V_2, \forall e_{ij} \in E_1\label{f2:c4}\\
    \text{with}\quad
    &y_{i,k} \in \{0, 1\} \quad \forall (u_i, v_k) \in V_1 \times V_2\label{f2:d1}\\
    &z_{ij,kl} \in \{0, 1\} \quad \forall (e_{ij}, e_{kl}) \in E_1 \times E_2\label{f2:d2}
  \end{align}
\end{subequations}
\label{model:F2}
\end{model}
$\gamma$ is not a function of $y$ and $z$. It does not impact the minimization problem. However, $\gamma$ is mandatory to obtain the GED value (i.e. $d_{min}(G_1,G_2)$ from Problem \ref{prob:ged}). The topological constraints \eqref{eq:topology_1} and \eqref{eq:topology_2} are expressed in another way and are replaced by the constraints \eqref{f2:c3} and \eqref{f2:c4}.





\section{Proposal on the unification of the two problems}
In this paragraph, we propose to draw a relation between the graph matching and graph edit distance problems. Especially, we create a link between both problems through a change of similarity functions. Our proposal can be stated as follows:
\begin{proposition}
\label{proposition:maxminthesamif}
GM and GED problems are equivalent in terms of solutions under a reformulation of the similarity function $s'(u_i \to v_k)= - \left(  c(u_i \to v_k) - c(u_i \to \epsilon ) - c(\epsilon \to v_k)  \right) $ and  $s'(e_{ij} \to e_{kl})= - \left(  c(e_{ij} \to e_{kl}) - c(e_{ij} \to \epsilon ) - c(\epsilon \to e_{kl})  \right)$
\end{proposition}
To intuitively demonstrate the exactness of the proposition, we proceed as follows :
\begin{enumerate}
    \item We start from the GED problem expressed by model F2 (see Model \ref{model:F2}). 
    \item We link the similarity function $s$ with the cost function $c$ thanks to a new similarity function $s'$.
    \item With this similarity function $s'$, we show that F2 turns to be a maximization problem and we call this new model F2'.
    \item F2' is modified by switching from a linear to a quadratic model called GMM'. 
    \item GMM' is identical to GMM. It is sufficient to show that both models express the same problem, that is to say, the graph matching problem.
\end{enumerate}
\begin{proof}
\begin{enumerate}
\item By setting  $d(u_i \to v_k)=  \Bigl( c(u_i \to v_k) - c(u_i \to \epsilon ) - c(\epsilon \to v_k)  \Bigr)$ and $d(e_{ij} \to e_{kl})=  \Bigl( c(e_{ij} \to e_{kl}) - c(e_{ij} \to \epsilon ) - c(\epsilon \to e_{kl})  \Bigr)$, we can rewrite the objective function of F2 as follows : 
\begin{equation}
 \begin{aligned}
  C'({y,z})=&\gamma+ \sum_{u_i \in V_1}\sum_{v_k \in V_2} d(u_i \to v_k)  \cdot y_{u_i,v_k} \\
  &+ \sum_{e_{ij} \in E_1}\sum_{e_{kl} \in E_2} d(e_{ij} \to e_{kl}) \cdot z_{ij,kl} \\
  \text{with } \gamma =& \sum_{u_i \in V_1} c(u_i \to \epsilon) + \sum_{v_k \in V_2} c(\epsilon \to v_k) \\
  &+ \sum_{e_{ij} \in E_1} c(e_{ij} \to \epsilon) + \sum_{e_{kl} \in E_2} c(\epsilon \to e_{kl})
\end{aligned}
\label{eq:objective4}
\end{equation}

\item $\gamma$ does not depend on variables so it does not impact the optimization problem. Therefore $\gamma$ can be removed.
\item By setting  $s'(u_i \to v_k)= -d(u_i \to v_k)= -\Bigl(c(u_i \to v_k) - c(u_i \to \epsilon ) - c(\epsilon \to v_k)  \Bigr)$ and similarly, $s'(e_{ij} \to e_{kl})= -d(e_{ij} \to e_{kl})$, we can rewrite the objective function $C'$ of the model F2 to obtain $S'$.

\begin{equation}
 \begin{aligned}
  S'({y,z})=& \sum_{u_i \in V_1}\sum_{v_k \in V_2} s'(u_i \to v_k)  \cdot y_{i,k}\\
  &+ \sum_{e_{ij} \in E_1}\sum_{e_{kl} \in E_2} s'(e_{ij} \to e_{kl}) \cdot z_{ij,kl}
\end{aligned}
\label{eq:objective3}
\end{equation}

\item In a general way, minimizing $f(x)$ is equivalent to maximize -$f(x)$. So, minimizing $C'$ is equivalent to maximize $S'$.

\item The linear objective function $S'$ can be turned into a quadratic function by removing variables $z$ and replacing them by product of $y$ variables. 

\begin{equation}
 \begin{aligned}
  S''(y)=& \sum_{u_i \in V_1}\sum_{v_k \in V_2} s'(u_i \to v_k)  \cdot y_{i,k} \\
  &+ \sum_{e_{ij} \in E_1}\sum_{e_{kl} \in E_2} s'(e_{ij} \to e_{kl}) \cdot y_{i,k} \cdot y_{j,l} \\
\end{aligned}
\label{eq:objective4}
\end{equation}

\item Topological constraints (Equations \eqref{f2:c3} and \eqref{f2:c4}) in F2 are not necessary anymore and they can be removed. The product of $y_{i,k}$ and $y_{j,l}$ is enough to ensure that an edge $e_{ij} \in E_1$ can be matched to an edge $e_{kl} \in E_2$ only if the head vertices $u_i \in V_1$ and $v_k \in V_2$, on the one hand, and if the tail vertices $u_j \in V_1$ and $v_l \in V_2$, on the other hand, are respectively matched.

\item We obtain the new model named GMM': 

\begin{model}{GMM'}
\begin{subequations}
    \begin{align}
    y^* =& \underset{y} {\mathrm{argmax}} \quad S''(y) \label{qap:o} \\
    \text{subject to}\quad
    &\sum_{u_i \in V_1} y_{i,k} \leq 1 \quad \forall v_k \in V_2\label{qap:c2}\\
     &\sum_{v_k \in V_2} y_{i,k} \leq 1 \quad \forall u_i \in V_1\label{qap:c1}\\
    \text{with}\quad
    &y_{i,k} \in \{0, 1\} \quad \forall (u_i, v_k) \in V_1 \times V_2\label{qap:d1}
  \end{align}
\end{subequations}
\label{model:QAP2}
\end{model}


\item Model GMM' = Model GMM. This was to be demonstrated. Proposition \ref{proposition:maxminthesamif} is right.
\end{enumerate}
\end{proof}

\begin{figure}
    \centering
    \includegraphics[width=.5\textwidth]{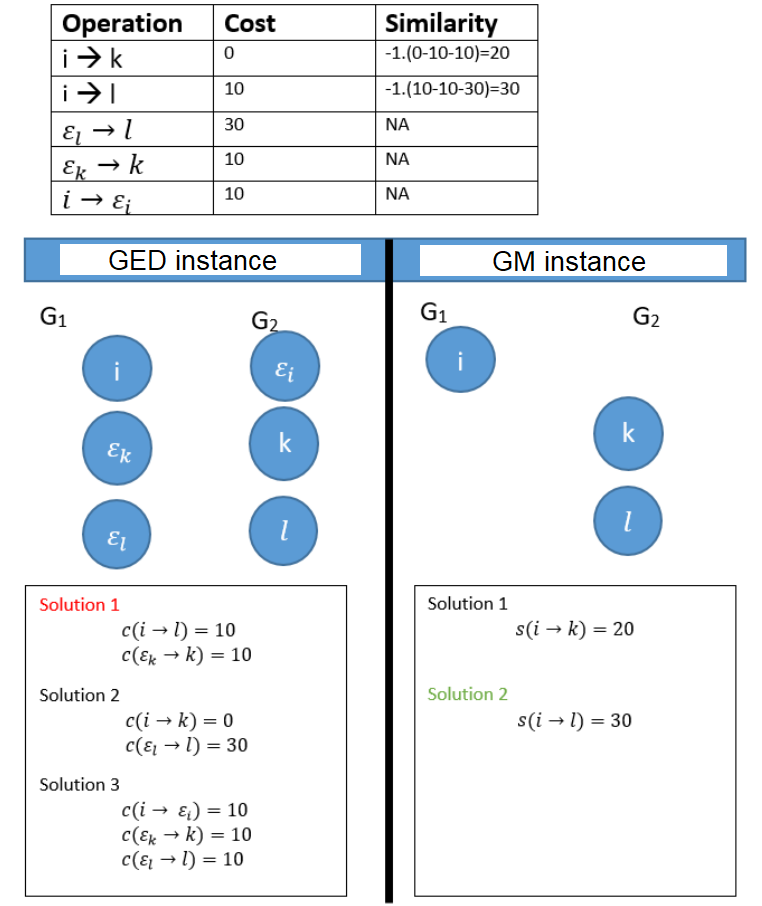}
    \caption{A comparison of the graph matching and GED problems when the similarity function  $s'(i \to k)= - \{ c(i \to k) - c(i \to \epsilon ) - c(\epsilon \to k)  \}$}
    \label{fig:sgmgmproposition3}
\end{figure}

Under the condition of Proposition \ref{proposition:maxminthesamif}, the optimal assignment obtains when solving the graph matching problem can be used to reconstruct an optimal solution of the GED problem. An instance of GED and an instance of GM are presented in Figure \ref{fig:sgmgmproposition3}. Solutions of the GED instance are presented with respect to the cost function $c$ while the graph matching solutions are presented with respect to the similarity function $s'$. The optimal matching of both instances are the same.

\color{black} Model GMM' has $|V_1| . |V_2|$ variables and $ |V_1| + |V_2|$ constraints. Similarity functions can be represented by a similarity matrix $K$ of size is $|V_1| . |V_2| \times |V_1| . |V_2|$.\color{black} 

Proposition \ref{proposition:maxminthesamif} is a first attempt toward the unification of two communities working respectively on GED and GM problems. All the methods solving the graph matching problem can be used to solve the graph edit distance problem under a specific similarity function $s'(u_i \to v_k)= - \Bigr( c(u_i \to v_k) - c(u_i \to \epsilon ) - c(\epsilon \to v_k)  \Bigr)$.

\color{black} \section{Experiments}
In this section, we show the results of our numerical experiments to validate our proposal that the model GMM' can model the GED problem if $s'(i \to k)= - \{ c(i \to k) - c(i \to \epsilon ) - c(\epsilon \to k)  \}$.
We based our protocol on the ICPR GED contest\footnote{ \url{https://gdc2016.greyc.fr/} \citep{ABUAISHEH201796}}. 
Among the data sets available, we chose the GREC data set for two reasons. First, graphs sizes range from 5 to 20 nodes and these sizes are amendable to compute optimal solutions. Second, the GREC cost function, defined in the contest, is complex enough to cover a large range of matching cases. This cost function is not a constant value and includes euclidean distances between point coordinates. The reader is redirected to \citep{ABUAISHEH201796} for the full definition of the cost function. From the GREC database, we chose the subset of graphs called "MIXED" because it holds 10 graphs of various sizes. We computed all the pairwise comparisons to obtain 100 solutions. We compared the optimal solutions obtained by our Model GMM' and the optimal solutions found by the straightforward ILP formulation called F1 \citep{DBLP:journals/pr/LerougeARHA17}. We computed the average difference between the GED values and the objective function values of our model GMM'. \textbf{The average difference is exactly equal to zero. This result corroborates our theoretical statement.} Detailed results and codes can be found on the website \url{https://sites.google.com/view/a-single-model-for-ged-and-gm}.

\color{black}
\section{Conclusion}
In this paper, an equivalence between graph matching and graph edit distance problems was proven under a reformulation of the similarity functions between nodes and edges. These functions should take into account explicitly the deletion and insertion costs. That's the major difference between GM and GED problems. In the GED problem, costs to delete or to insert vertices or edges are explicitly introduced in the error model. On the other hand, deletion costs are implicitly set to a specific value (that is to say 0) in the GM problem.
\color{black}Many learning methods aim at learning edit costs \citep{SERRATOSA2020115,MARTINEAU202068} or matching similarities \citep{deeplearningofgm,learningm}. Learned matching similarities may include implicitly deletion and insertion costs. Does it help the learning algorithm to learn separately insertion and deletion costs? That is still an open question. However, with this paper, we stand for a rapprochement of the research communities that work on learning graph edit distance and learning graph matching because edit costs can be hidden in the learned similarities.
\color{black}

\bibliographystyle{model1-num-names}
\bibliography{ref}

\begin{thebibliography}{33}
\expandafter\ifx\csname natexlab\endcsname\relax\def\natexlab#1{#1}\fi
\providecommand{\url}[1]{\texttt{#1}}
\providecommand{\href}[2]{#2}
\providecommand{\path}[1]{#1}
\providecommand{\DOIprefix}{doi:}
\providecommand{\ArXivprefix}{arXiv:}
\providecommand{\URLprefix}{URL: }
\providecommand{\Pubmedprefix}{pmid:}
\providecommand{\doi}[1]{\href{http://dx.doi.org/#1}{\path{#1}}}
\providecommand{\Pubmed}[1]{\href{pmid:#1}{\path{#1}}}
\providecommand{\bibinfo}[2]{#2}
\ifx\xfnm\relax \def\xfnm[#1]{\unskip,\space#1}\fi
\bibitem[{Riesen(2015)}]{Riesen2015bouquin}
\bibinfo{author}{K.~Riesen}, \bibinfo{title}{Structural Pattern Recognition
  with Graph Edit Distance - Approximation Algorithms and Applications},
  Advances in Computer Vision and Pattern Recognition,
  \bibinfo{publisher}{Springer}, \bibinfo{year}{2015}.
\bibitem[{Das and Lee(2018)}]{DAS201880}
\bibinfo{author}{D.~Das}, \bibinfo{author}{C.~G. Lee},
\newblock \bibinfo{title}{Sample-to-sample correspondence for unsupervised
  domain adaptation},
\newblock \bibinfo{journal}{Engineering Applications of Artificial
  Intelligence} \bibinfo{volume}{73} (\bibinfo{year}{2018}) \bibinfo{pages}{80
  -- 91}.
\bibitem[{Swoboda et~al.(2017)Swoboda, Rother, Abu~Alhaija, Kainmuller, and
  Savchynskyy}]{Swoboda_2017_CVPR}
\bibinfo{author}{P.~Swoboda}, \bibinfo{author}{C.~Rother},
  \bibinfo{author}{H.~Abu~Alhaija}, \bibinfo{author}{D.~Kainmuller},
  \bibinfo{author}{B.~Savchynskyy},
\newblock \bibinfo{title}{A study of lagrangean decompositions and dual ascent
  solvers for graph matching},
\newblock in: \bibinfo{booktitle}{CVPR}, \bibinfo{year}{2017}.
\bibitem[{Garey and Johnson(1979)}]{GMcomplexity}
\bibinfo{author}{M.~R. Garey}, \bibinfo{author}{D.~S. Johnson},
  \bibinfo{title}{Computers and Intractability; A Guide to the Theory of
  NP-Completeness}, \bibinfo{publisher}{W. H. Freeman Co.},
  \bibinfo{address}{USA}, \bibinfo{year}{1979}.
\bibitem[{Tsai et~al.(1979)Tsai, Member, and Fu}]{Tsai1979}
\bibinfo{author}{W.-h. Tsai}, \bibinfo{author}{S.~Member},
  \bibinfo{author}{K.-s. Fu},
\newblock \bibinfo{title}{{Pattern Deformational Model and Bayes
  Error-Correcting Recognition System}},
\newblock \bibinfo{journal}{IEEE Transactions on Systems, Man, and Cybernetics}
  \bibinfo{volume}{9} (\bibinfo{year}{1979}) \bibinfo{pages}{745--756}.
\bibitem[{Zeng et~al.(2009)Zeng, Tung, Wang, Feng, and
  Zhou}]{DBLP:journals/pvldb/ZengTWFZ09}
\bibinfo{author}{Z.~Zeng}, \bibinfo{author}{A.~K.~H. Tung},
  \bibinfo{author}{J.~Wang}, \bibinfo{author}{J.~Feng},
  \bibinfo{author}{L.~Zhou},
\newblock \bibinfo{title}{Comparing stars: On approximating graph edit
  distance},
\newblock \bibinfo{journal}{PVLDB} \bibinfo{volume}{2} (\bibinfo{year}{2009})
  \bibinfo{pages}{25--36}.
\bibitem[{Bougleux et~al.(2017)Bougleux, Brun, Carletti, Foggia, Gauzere, and
  Vento}]{bougleuxQAP}
\bibinfo{author}{S.~Bougleux}, \bibinfo{author}{L.~Brun},
  \bibinfo{author}{V.~Carletti}, \bibinfo{author}{P.~Foggia},
  \bibinfo{author}{B.~Gauzere}, \bibinfo{author}{M.~Vento},
\newblock \bibinfo{title}{Graph edit distance as a quadratic assignment
  problem},
\newblock \bibinfo{journal}{Pattern Recognition Letters} \bibinfo{volume}{87}
  (\bibinfo{year}{2017}) \bibinfo{pages}{38 -- 46}. \bibinfo{note}{Advances in
  Graph-based Pattern Recognition}.
\bibitem[{Cho et~al.(2013)Cho, Alahari, and Ponce}]{DBLP:conf/iccv/ChoAP13}
\bibinfo{author}{M.~Cho}, \bibinfo{author}{K.~Alahari},
  \bibinfo{author}{J.~Ponce},
\newblock \bibinfo{title}{Learning graphs to match},
\newblock in: \bibinfo{booktitle}{ICCV}, \bibinfo{year}{2013}, pp.
  \bibinfo{pages}{25--32}.
\bibitem[{Torresani et~al.(2013)Torresani, Kolmogorov, and
  Rother}]{messagepassingdualdecomposition}
\bibinfo{author}{L.~Torresani}, \bibinfo{author}{V.~Kolmogorov},
  \bibinfo{author}{C.~Rother},
\newblock \bibinfo{title}{A dual decomposition approach to feature
  correspondence},
\newblock \bibinfo{journal}{TPAMI} \bibinfo{volume}{35} (\bibinfo{year}{2013})
  \bibinfo{pages}{259--271}.
\bibitem[{Liu and Qiao(2014)}]{gnccp}
\bibinfo{author}{Z.~Liu}, \bibinfo{author}{H.~Qiao},
\newblock \bibinfo{title}{Gnccp—graduated nonconvexityand concavity
  procedure},
\newblock \bibinfo{journal}{TPAMI} \bibinfo{volume}{36} (\bibinfo{year}{2014})
  \bibinfo{pages}{1258--1267}.
\bibitem[{Schellewald and Schn{\"o}rr(2005)}]{pbmatching}
\bibinfo{author}{C.~Schellewald}, \bibinfo{author}{C.~Schn{\"o}rr},
\newblock \bibinfo{title}{Probabilistic subgraph matching based on convex
  relaxation},
\newblock in: \bibinfo{editor}{A.~Rangarajan}, \bibinfo{editor}{B.~Vemuri},
  \bibinfo{editor}{A.~L. Yuille} (Eds.), \bibinfo{booktitle}{Energy
  Minimization Methods in Computer Vision and Pattern Recognition},
  \bibinfo{publisher}{Springer Berlin Heidelberg}, \bibinfo{address}{Berlin,
  Heidelberg}, \bibinfo{year}{2005}, pp. \bibinfo{pages}{171--186}.
\bibitem[{Bazaraa and Sherali(1982)}]{Bazaraa1982}
\bibinfo{author}{M.~S. Bazaraa}, \bibinfo{author}{H.~D. Sherali},
\newblock \bibinfo{title}{On the use of exact and heuristic cutting plane
  methods for the quadratic assignment problem},
\newblock \bibinfo{journal}{Journal of the Operational Research Society}
  \bibinfo{volume}{33} (\bibinfo{year}{1982}) \bibinfo{pages}{991--1003}.
\bibitem[{Gold and Rangarajan(1996)}]{graduatedassignment}
\bibinfo{author}{S.~Gold}, \bibinfo{author}{A.~Rangarajan},
\newblock \bibinfo{title}{A graduated assignment algorithm for graph matching},
\newblock \bibinfo{journal}{TPAMI} \bibinfo{volume}{18} (\bibinfo{year}{1996})
  \bibinfo{pages}{377--388}.
\bibitem[{Leordeanu et~al.(2009)Leordeanu, Hebert, and Sukthankar}]{IPFP}
\bibinfo{author}{M.~Leordeanu}, \bibinfo{author}{M.~Hebert},
  \bibinfo{author}{R.~Sukthankar},
\newblock \bibinfo{title}{An integer projected fixed point method for graph
  matching and map inference},
\newblock in: \bibinfo{editor}{Y.~Bengio}, \bibinfo{editor}{D.~Schuurmans},
  \bibinfo{editor}{J.~D. Lafferty}, \bibinfo{editor}{C.~K.~I. Williams},
  \bibinfo{editor}{A.~Culotta} (Eds.), \bibinfo{booktitle}{NIPS},
  \bibinfo{publisher}{Curran Associates, Inc.}, \bibinfo{year}{2009}, pp.
  \bibinfo{pages}{1114--1122}.
\bibitem[{Cour et~al.(2007)Cour, Srinivasan, and Shi}]{SMACGM}
\bibinfo{author}{T.~Cour}, \bibinfo{author}{P.~Srinivasan},
  \bibinfo{author}{J.~Shi},
\newblock \bibinfo{title}{Balanced graph matching},
\newblock in: \bibinfo{editor}{B.~Sch\"{o}lkopf}, \bibinfo{editor}{J.~C.
  Platt}, \bibinfo{editor}{T.~Hoffman} (Eds.), \bibinfo{booktitle}{NIPS},
  \bibinfo{publisher}{MIT Press}, \bibinfo{year}{2007}, pp.
  \bibinfo{pages}{313--320}.
\bibitem[{Leordeanu and Hebert(2005)}]{spectralmatching}
\bibinfo{author}{M.~Leordeanu}, \bibinfo{author}{M.~Hebert},
\newblock \bibinfo{title}{A spectral technique for correspondence problems
  using pairwise constraints},
\newblock in: \bibinfo{booktitle}{ICCV}, volume~\bibinfo{volume}{2},
  \bibinfo{year}{2005}, pp. \bibinfo{pages}{1482--1489 Vol. 2}.
\bibitem[{Cho et~al.(2010)Cho, Lee, and Lee}]{reweightedgm}
\bibinfo{author}{M.~Cho}, \bibinfo{author}{J.~Lee}, \bibinfo{author}{K.~M.
  Lee},
\newblock \bibinfo{title}{Reweighted random walks for graph matching},
\newblock in: \bibinfo{editor}{K.~Daniilidis}, \bibinfo{editor}{P.~Maragos},
  \bibinfo{editor}{N.~Paragios} (Eds.), \bibinfo{booktitle}{ECCV},
  \bibinfo{publisher}{Springer Berlin Heidelberg}, \bibinfo{address}{Berlin,
  Heidelberg}, \bibinfo{year}{2010}, pp. \bibinfo{pages}{492--505}.
\bibitem[{Riesen et~al.(2007)Riesen, Fankhauser, and
  Bunke}]{DBLP:conf/mlg/RiesenFB07}
\bibinfo{author}{K.~Riesen}, \bibinfo{author}{S.~Fankhauser},
  \bibinfo{author}{H.~Bunke},
\newblock \bibinfo{title}{Speeding up graph edit distance computation with a
  bipartite heuristic},
\newblock in: \bibinfo{booktitle}{Mining and Learning with Graphs,
  Proceedings}, \bibinfo{year}{2007}.
\bibitem[{Abu{-}Aisheh et~al.(2015)Abu{-}Aisheh, Raveaux, Ramel, and
  Martineau}]{DBLP:conf/icpram/Abu-AishehRRM15}
\bibinfo{author}{Z.~Abu{-}Aisheh}, \bibinfo{author}{R.~Raveaux},
  \bibinfo{author}{J.~Ramel}, \bibinfo{author}{P.~Martineau},
\newblock \bibinfo{title}{An exact graph edit distance algorithm for solving
  pattern recognition problems},
\newblock in: \bibinfo{booktitle}{{ICPRAM}}, \bibinfo{year}{2015}, pp.
  \bibinfo{pages}{271--278}.
\bibitem[{Justice and Hero(2006)}]{justiceheroged2006}
\bibinfo{author}{D.~Justice}, \bibinfo{author}{A.~Hero},
\newblock \bibinfo{title}{A binary linear programming formulation of the graph
  edit distance},
\newblock \bibinfo{journal}{TPAMI} \bibinfo{volume}{28} (\bibinfo{year}{2006})
  \bibinfo{pages}{1200--1214}.
\bibitem[{Lerouge et~al.(2017)Lerouge, Abu{-}Aisheh, Raveaux, H{\'{e}}roux, and
  Adam}]{DBLP:journals/pr/LerougeARHA17}
\bibinfo{author}{J.~Lerouge}, \bibinfo{author}{Z.~Abu{-}Aisheh},
  \bibinfo{author}{R.~Raveaux}, \bibinfo{author}{P.~H{\'{e}}roux},
  \bibinfo{author}{S.~Adam},
\newblock \bibinfo{title}{New binary linear programming formulation to compute
  the graph edit distance},
\newblock \bibinfo{journal}{Pattern Recognition} \bibinfo{volume}{72}
  (\bibinfo{year}{2017}) \bibinfo{pages}{254--265}.
\bibitem[{Bougleux et~al.(2017)Bougleux, Ga{\"u}z{\`e}re, and
  Brun}]{bougleuxlsap}
\bibinfo{author}{S.~Bougleux}, \bibinfo{author}{B.~Ga{\"u}z{\`e}re},
  \bibinfo{author}{L.~Brun},
\newblock \bibinfo{title}{A hungarian algorithm for error-correcting graph
  matching},
\newblock in: \bibinfo{editor}{P.~Foggia}, \bibinfo{editor}{C.-L. Liu},
  \bibinfo{editor}{M.~Vento} (Eds.), \bibinfo{booktitle}{Graph-Based
  Representations in Pattern Recognition}, \bibinfo{publisher}{Springer
  International Publishing}, \bibinfo{address}{Cham}, \bibinfo{year}{2017}, pp.
  \bibinfo{pages}{118--127}.
\bibitem[{Serratosa(2015)}]{DBLP:journals/ivc/Serratosa15}
\bibinfo{author}{F.~Serratosa},
\newblock \bibinfo{title}{Computation of graph edit distance: Reasoning about
  optimality and speed-up},
\newblock \bibinfo{journal}{Image Vision Comput.} \bibinfo{volume}{40}
  (\bibinfo{year}{2015}) \bibinfo{pages}{38--48}.
\bibitem[{Riesen and Bunke(2009)}]{DBLP:journals/ivc/RiesenB09}
\bibinfo{author}{K.~Riesen}, \bibinfo{author}{H.~Bunke},
\newblock \bibinfo{title}{Approximate graph edit distance computation by means
  of bipartite graph matching},
\newblock \bibinfo{journal}{Image Vision Comput.} \bibinfo{volume}{27}
  (\bibinfo{year}{2009}) \bibinfo{pages}{950--959}.
\bibitem[{Neuhaus and Bunke.(2007)}]{Neuhaus2007}
\bibinfo{author}{M.~Neuhaus}, \bibinfo{author}{H.~Bunke.},
\newblock \bibinfo{title}{Bridging the gap between graph edit distance and
  kernel machines.},
\newblock \bibinfo{journal}{Machine Perception and Artificial Intelligence.}
  \bibinfo{volume}{68} (\bibinfo{year}{2007}) \bibinfo{pages}{17--61}.
\bibitem[{Bunke(1997)}]{DBLP:journals/prl/Bunke97}
\bibinfo{author}{H.~Bunke},
\newblock \bibinfo{title}{On a relation between graph edit distance and maximum
  common subgraph},
\newblock \bibinfo{journal}{Pattern Recognition Letters} \bibinfo{volume}{18}
  (\bibinfo{year}{1997}) \bibinfo{pages}{689--694}.
\bibitem[{Bunke(1999)}]{DBLP:journals/pami/Bunke99}
\bibinfo{author}{H.~Bunke},
\newblock \bibinfo{title}{Error correcting graph matching: On the influence of
  the underlying cost function},
\newblock \bibinfo{journal}{TPAMI} \bibinfo{volume}{21} (\bibinfo{year}{1999})
  \bibinfo{pages}{917--922}.
\bibitem[{Brun et~al.(2012)Brun, Ga{\"u}z{\`e}re, and
  Fourey}]{brun:hal-00714879}
\bibinfo{author}{L.~Brun}, \bibinfo{author}{B.~Ga{\"u}z{\`e}re},
  \bibinfo{author}{S.~Fourey}, \bibinfo{title}{{Relationships between Graph
  Edit Distance and Maximal Common Unlabeled Subgraph}},
  \bibinfo{type}{Technical Report}, \bibinfo{year}{2012}.
\bibitem[{Abu-Aisheh et~al.(2017)Abu-Aisheh, Gauzere, Bougleux, and
  et~al}]{ABUAISHEH201796}
\bibinfo{author}{Z.~Abu-Aisheh}, \bibinfo{author}{B.~Gauzere},
  \bibinfo{author}{S.~Bougleux}, \bibinfo{author}{et~al},
\newblock \bibinfo{title}{Graph edit distance contest: Results and future
  challenges},
\newblock \bibinfo{journal}{Pattern Recognition Letters} \bibinfo{volume}{100}
  (\bibinfo{year}{2017}) \bibinfo{pages}{96 -- 103}.
\bibitem[{Serratosa(2020)}]{SERRATOSA2020115}
\bibinfo{author}{F.~Serratosa},
\newblock \bibinfo{title}{A general model to define the substitution, insertion
  and deletion graph edit costs based on an embedded space},
\newblock \bibinfo{journal}{Pattern Recognition Letters} \bibinfo{volume}{138}
  (\bibinfo{year}{2020}) \bibinfo{pages}{115 -- 122}.
\bibitem[{Martineau et~al.(2020)Martineau, Raveaux, Conte, and
  Venturini}]{MARTINEAU202068}
\bibinfo{author}{M.~Martineau}, \bibinfo{author}{R.~Raveaux},
  \bibinfo{author}{D.~Conte}, \bibinfo{author}{G.~Venturini},
\newblock \bibinfo{title}{Learning error-correcting graph matching with a
  multiclass neural network},
\newblock \bibinfo{journal}{Pattern Recognition Letters} \bibinfo{volume}{134}
  (\bibinfo{year}{2020}) \bibinfo{pages}{68 -- 76}.
\bibitem[{{Zanfir} and {Sminchisescu}(2018)}]{deeplearningofgm}
\bibinfo{author}{A.~{Zanfir}}, \bibinfo{author}{C.~{Sminchisescu}},
\newblock \bibinfo{title}{Deep learning of graph matching},
\newblock in: \bibinfo{booktitle}{2018 IEEE/CVF Conference on Computer Vision
  and Pattern Recognition}, \bibinfo{year}{2018}, pp.
  \bibinfo{pages}{2684--2693}.
\bibitem[{{Caetano} et~al.(2007){Caetano}, {Li Cheng}, {Le}, and
  {Smola}}]{learningm}
\bibinfo{author}{T.~S. {Caetano}}, \bibinfo{author}{{Li Cheng}},
  \bibinfo{author}{Q.~V. {Le}}, \bibinfo{author}{A.~J. {Smola}},
\newblock \bibinfo{title}{Learning graph matching},
\newblock in: \bibinfo{booktitle}{2007 IEEE 11th International Conference on
  Computer Vision}, \bibinfo{year}{2007}, pp. \bibinfo{pages}{1--8}.

\end{thebibliography}







\end{document}